\documentclass[twocolumn,superscriptaddress,preprintnumbers,amsmath,amssymb,nofootinbib]{revtex4}

\usepackage{amsmath}
\usepackage{amssymb}
\usepackage{slashed}
\usepackage{graphicx}
\usepackage{graphics}
\usepackage{epsfig}
\usepackage{color}
\usepackage{tabularx}
\usepackage{hyperref}
\newcommand{\be}{\begin{equation}}
\newcommand{\ee}{\end{equation}}
\newcommand{\bea}{\begin{eqnarray}}
\newcommand{\eea}{\end{eqnarray}}

\begin{document}
\title{Consistency of matter models with asymptotically safe quantum gravity}

\author{Pietro Don\`a}
\email[]{pietro.dona@sissa.it}
\affiliation{International School for Advanced Studies, via Bonomea 265, 34136 Trieste, Italy\\
and INFN, Sezione di Trieste}

\author{Astrid Eichhorn\footnote{Based on a talk given by A. Eichhorn at Theory Canada 9.}}
\email[]{aeichhorn@perimeterinstitute.ca} 
\affiliation{Perimeter Institute for Theoretical Physics, 31 Caroline Street N, Waterloo, N2L 2Y5, Ontario, Canada}

\author{Roberto Percacci}
\email[]{percacci@sissa.it} 
\affiliation{International School for Advanced Studies, via Bonomea 265, 34136 Trieste, Italy\\
and INFN, Sezione di Trieste}

\begin{abstract}
We discuss the compatibility of quantum gravity with dynamical matter degrees of freedom. 
Specifically, we present bounds we obtained in \cite{Dona:2013qba} on the allowed number and type of matter fields within asymptotically safe quantum gravity. As a novel result, we show bounds on the allowed number of spin-3/2 (Rarita-Schwinger) fields, e.g., the gravitino.
These bounds, obtained within truncated Renormalization Group flows, indicate the compatibility of asymptotic safety with the matter fields of the standard model. Further, they suggest that extensions of the matter content of the standard model are severely restricted in asymptotic safety. This means that searches for new particles at colliders could provide experimental tests for this particular approach to quantum gravity.
\end{abstract}

\maketitle

\section{Introduction: Experimental search for quantum gravity}
Research on quantum gravity is theoretical to an overwhelmingly large extent. Experimental guidance to gain a better understanding of the fundamental quantum nature of spacetime is very hard to come by, mostly because the ``natural" scale of quantum gravity, set by the Planck mass, $M_{\rm Planck} \approx 10^{19}\, \rm GeV$, is much higher than scales that can be accessed, e.g., in collider experiments. Here, our main goal will be to show that nevertheless ongoing searches for new particles at colliders could provide tests of quantum gravity models.
We will achieve our goal by showing that, given a particular candidate for a quantum gravity model, namely asymptotic safety, it is \emph{not} possible to include arbitrary dynamical matter degrees of freedom without destroying the consistency of the theory, at least within the approximation considered in \cite{Dona:2013qba}: There are bounds on the allowed number of fermions and scalars within asymptotically safe quantum gravity.

The main idea relies on the fact that the ultraviolet, i.e., microscopic, behavior of quantum field theories, such as asymptotically safe quantum gravity, depends on their field content:
Only for specific choices of the field content can we construct a quantum field theory that is ultraviolet (UV) complete, i.e., a theory where we can consistently study arbitrarily high momenta. 
For instance, Yang-Mills theory in four dimensions is well-behaved in the ultraviolet, where it becomes asymptotically free. In more technical terms, it approaches a noninteracting Renormalization Group (RG) fixed point. At such a fixed point, a quantum theory becomes scale-free, which allows us to study the limit of arbitrarily high momenta, without running into divergences. The fixed point changes its nature radically, if we couple too many quark flavors to Yang-Mills theory: Beyond a critical value, the fixed point becomes ultraviolet-repulsive, and we can no longer study the limit of arbitrarily high momenta. Thus, coupling too many matter degrees of freedom destroys the ultraviolet completion. In quantum gravity, indications for a similar effect were derived by us in \cite{Dona:2013qba}:  Within an approximation of the Renormalization Group flow, we found indications that quantum gravity cannot become asymptotically safe, if too many matter fields exist. These allow us to set up consistency tests of quantum gravity: If  one could show that the model fails to work when coupled to the standard model fields, then it would already be ruled out experimentally. Further, we showed in \cite{Dona:2013qba} that the discovery of new particles, e.g., at the LHC, could potentially provide a new way to experimentally rule out asymptotically safe quantum gravity.

\section{Renormalization Group flow in gravity and asymptotic safety}
Our universe contains gravitational as well as matter degrees of freedom. Here, we will assume that all of these are fundamental, i.e., they should be included in a microscopic quantum description. Our goal will be to find a consistent dynamics for matter and quantum spacetime within the quantum field theory framework. As the Newton coupling has negative mass-dimensionality, the corresponding quantum field theory cannot be asymptotically free: Its negative canonical dimensionality implies that the Newton coupling will grow towards high momenta. This does not rule out the possibility of finding a quantum field theory of gravity:
Weinberg has proposed \cite{Weinberg:1980gg} a generalization of asymptotic freedom, where a quantum field theory can be extended to arbitrarily high momentum scales, if it becomes asymptotically safe. Within this scenario, the running, momentum-scale $k$ dependent couplings approach constant, finite values in the ultraviolet (UV). Here, the couplings are taken to be dimensionless, i.e., their canonical scaling dimension is absorbed into an appropriate rescaling by powers of $k$. This allows to discover Renormalization Group fixed points, at which the theory becomes scale-free, and the limit $k \rightarrow \infty$ can be taken safely. This yields a microscopic dynamics, defined by the action at the fixed point. The main difference between asymptotic freedom and asymptotic safety lies in the fact, that in the latter case the fixed point is an interacting one, i.e., (a subset of) the couplings remains finite. On the technical side, the search for asymptotic safety requires a nonperturbative method to discover the fixed point, which can -- in principle -- lie at large values of the couplings. The notion of asymptotic safety is not restricted to gravity, but is also viable for other quantum field theories \cite{Braun:2010tt,Litim:2014uca}. In particular, asymptotic safety as a UV completion for our universe would require that all gravitational and matter couplings approach a fixed point.
The main physical assumptions underlying asymptotically safe quantum gravity are that there is no naive fundamental cutoff, i.e., the Planck scale does \emph{not} play the role of a naive discretization scale (see, however \cite{Percacci:2010af}). Instead, a ``quantum-gravity scale", at which the properties of the theory change radically, is generated dynamically, very much like the scale $\Lambda_{\rm QCD}$ in QCD. A second assumption is that the metric carries the microscopic degrees of freedom in gravity. These two assumptions lead to a mathematical formulation of the theory that differs considerably from most other research on quantum gravity.

In order to study the asymptotic safety scenario, we have to understand the Renormalization Group flow in gravity. It describes how effective theories, encoding dynamics for fields at a momentum scale $k$, change, when the scale $k$ changes. $k$ can intuitively be understood as the ``resolution" scale of the effective theory. To transition from $k$ to, e.g., $k - \delta k$, quantum fluctuations in the momentum shell of width $\delta k$ below $k$ have to be integrated out in the path integral.  In this step, quantum fluctuations generate all possible interactions compatible with the field content and symmetries of the theory. Accordingly, the RG flow is set up within an infinite dimensional space of all possible couplings, which is called theory space. For instance, in the case of gravity, theory space contains the Newton coupling, the cosmological constant, and couplings of all possible higher-curvature and higher-derivative operators.
The RG flow provides trajectories in this space, parameterized by $k$, which connect the effective dynamics at all possible scales. We call a theory fundamental, if $k$ can be extended to arbitrarily large values, i.e., into the microscopic regime, and to $k \rightarrow 0$, i.e., arbitrarily small momenta, without encountering divergences in any of the couplings. In contrast, a trajectory which hits a divergence at a scale $k'$ is called effective, and the scale $k'$ should be interpreted as the scale of ``new physics". For instance, if the fields that have been used in the description are composite, then $k'$ would signal the scale of compositeness. In the case of gravity, we will be looking for trajectories which can be extended up to $k \rightarrow \infty$, where the interacting RG fixed point will ``protect" the theory from running into divergences.

\subsection{Predictivity in asymptotic safety}
In an infinite-dimensional space of couplings, the theory can remain predictive, if the fixed point is UV attractive in only finitely many directions: In that case, one can pick an RG trajectory with vanishing projection on the UV repulsive directions, i.e., a trajectory within the UV critical hypersurface of the fixed point. This ensures that the trajectory hits the fixed point in the far UV. Furthermore, it implies that only finitely many free parameters are left to be determined by comparison with experiment: The infrared values of couplings (or combinations thereof) that correspond to UV attractive eigendirections are free parameters, since any choice of their values is compatible with the fixed-point requirement.

More specifically, let us consider the linearized RG flow around a fixed point with coordinates $g_{j\, \ast}$ for the couplings $g_j$. Here, we have made a transition to dimensionless couplings $g_j = \bar{g}_j k^{-d_{g_j}}$, where $d_{g_j}$ is the canonical dimensionality of the coupling $\bar{g}_j$. Then,
\be
\beta_{g_j} =\sum_i \frac{\partial \beta_{g_j}}{\partial g_i} \Big|_{g_i =g_{i\, \ast}}\left(g_{i}- g_{i\, \ast} \right)+ \mathcal{O}\left(g_{i\, \ast}- g_{i} \right)^2.
\ee
The first term in the expansion vanishes because of the fixed-point requirement, $\beta_{g_j}(g_{i\, \ast})=0$.
The solution to this equation reads
\be
g_j(k) = g_{j\, \ast} + \sum_I C_I V_i^{I} \left( \frac{k}{k_0}\right)^{- \theta_I}.
\ee
Herein, $V_I$ are the eigenvectors of the stability matrix
\be
B_{ij} = \frac{\partial \beta_{g_i}}{\partial{g_j}} \Big|_{g_l = g_{l \, \ast}}, \label{stabm}
\ee
and $\theta_I$ are its eigenvalues, multiplied by an additional negative sign. The $C_I$ are constants of integration. In order to hit the fixed point $g_{j\, \ast}$ in the ultraviolet, one has to set $C_I =0$ for all those eigenvectors for which $\theta_I<0$, i.e., the irrelevant couplings. On the other hand, the $C_I$ are abitrary for those directions for which $\theta_I>0$. These are the free parameters of the theory. In the case of the Gau\ss{}ian, noninteracting fixed point, the $\theta_I$ correspond to the canonical dimensionality of the couplings. In the case of an interacting fixed point, the critical exponents $\theta_I$ receive corrections from nonvanishing interactions. The infrared values of couplings depend on the choice of the $C_I$. Accordingly, the free parameters of the theory can be fixed by comparison to experiment in the infrared.

To evalute the RG flow within the full theory space is extremely challenging in practice, and one thus resorts to evaluations of the flow within (finite-dimensional) subspaces, i.e., truncations of the full theory space. In the case of gravity, a variety of such truncations have yielded considerable evidence for the existence of an interacting fixed point \cite{Reuter:1996cp,Codello:2008vh,Benedetti:2009rx,Benedetti:2012dx,Dietz:2012ic,Falls:2013bv}, see also \cite{Reuter:2012id} for further references. 

\subsection{Setting a scale in quantum gravity}
Most evidence for asymptotic safety relies on the use of the Wetterich equation \cite{Wetterich:1993yh,Morris:1993qb} for the scale dependent effective action $\Gamma_k$, which includes the effect of quantum fluctuations with momenta higher than $k$. In quantum gravity, where spacetime itself fluctuates, it is more challenging than within quantum field theories on a flat background to set a scale and sort quantum fluctuations according to their momenta. Every metric configuration in the path integral comes with its own notion of scale, encoded in the Laplacian $-g^{\mu \nu}D_{\mu}D_{\nu}$. This multitude of possible scale-settings is due to diffeomorphism invariance, which implies background independence and puts quantum gravity on a conceptually very different footing from standard flat-background quantum field theories.
Here, the background field method \cite{Abbott:1980hw} allows to make progress: Picking a (fiducial) background metric $\bar{g}_{\mu \nu}$ facilitates a split of any configuration $g_{\mu \nu}$ on the same topology into
\be
g_{\mu \nu} = \bar{g}_{\mu \nu} + h_{\mu \nu}.
\ee
The path-integral over $g_{\mu \nu}$ can then be evaluated by integrating over configurations $h_{\mu \nu}$, which can have nonperturbatively large amplitudes.
The background metric $\bar{g}_{\mu \nu}$ is used to set a scale: Any configuration $h_{\mu \nu}$ can be decomposed into eigenfunctions of the Laplacian $-\bar{D}^2$ on the background. Contributions with eigenvalue $\lambda> k^2$ should be included in $\Gamma_k$, while those with lower eigenvalue are not yet integrated out. This is facilitated by a scale dependent, mass-like term that is added within the exponential in the partition function, and takes the form $h_{\mu \nu}R^{k\, \mu \nu \kappa \lambda}(-\bar{D}^2) h_{\kappa \lambda}$ (note that we work in a Euclidean setting here). We demand that $R_k(x) =0$ for $x/k^2>1$, thus ensuring that high-momentum quantum fluctuations are integrated out in the path integral. In contrast, $R_k(x)>0$ for $x/k^2<1$, such that low-momentum quantum fluctuations are surpressed and not yet integrated out. The background field method thus provides an answer to the apparent problem of setting a scale in quantum gravity. As a drawback, it implies a technical complication: The dependence of the regulator on the background metric and fluctuation field is such that these cannot be combined into a full metric $g_{\mu \nu}$. The same is true for the gauge-fixing term \footnote{Here, we apply a background-field dependent gauge-fixing of the form
\be
F_{\mu}= \frac{\sqrt{2}}{\sqrt{32 \pi \bar{G}}} \left(\bar{D}^{\nu}h_{\mu \nu}- \frac{1}{2} \bar{D}_{\mu}h^{\nu}_{\nu} \right).
\ee
The gauge-fixing and Faddeev-Popov ghost action take the form
\bea
S_{\rm gf}&=&\frac{1}{2} \int d^dx\, \sqrt{\bar{g}}\, \bar{g}^{\mu \nu}F_{\mu}F_{\nu},\\
S_{\rm ghost}&=& - \sqrt{2}Z_c \int d^dx\, \sqrt{\bar{g}} \bar{c}_{\mu} (\bar{D}^{\rho}\bar{g}^{\mu \kappa} g_{\kappa \nu}D_{\rho} \nonumber\\
&{}&+ \bar{D}^{\rho} \bar{g}^{\mu \kappa}g_{\rho \nu} D_{\kappa} - \bar{D}^{\mu} \bar{g}^{\rho \sigma} g_{\rho \nu}D_{\sigma}) c^{\nu}.
\eea
Here, we have included a scale-dependent wave function renormalization $Z_c$ for the ghosts, that was first studied in \cite{Eichhorn:2010tb,Groh:2010ta}.}. 
This entails that couplings pertaining to fluctuation fields and couplings pertaining to background operators run differently. Accordingly, when we decompose, e.g.,  $\sqrt{g}R$ into a term that is quadratic in the fluctuation, schematically $\sqrt{\bar{g}}h_{\mu \nu}\bar{D}^2 h_{\mu \nu}$ (and further tensor structures), and a background curvature term $\sqrt{\bar{g}} \bar{R}$, then the corresponding couplings have to be disentangled.

\subsection{The Wetterich equation for quantum gravity and matter}
To derive the beta functions, we will employ the Wetterich equation, that governs
the scale-dependence of the effective action $\Gamma_k$. M. Reuter pioneered its application to gravity in \cite{Reuter:1996cp}.  The Wetterich equation is formulated in terms of the dimensionless scale derivative $\partial_t = k \partial_k$ and reads
\be
\partial_t\Gamma_k=\frac{1}{2} {\rm Str}\left(\Gamma_k^{(2)}+R_k \right)^{-1} \partial_t R_k.
\ee
Herein, the $\rm Str$ stands for a summation over all fields, including a negative sign for Grassmann-valued fields, e.g., fermions and Faddeev-Popov ghost fields. Additionally, the trace stands for a summation/integration over the discrete/continuous eigenvalue spectrum of the regularized nonperturbative propagator $\left( \Gamma_k^{(2)} +R_k\right)^{-1}$. Herein, $\Gamma_k^{(2)}$ is the second functional derivative with respect to the fields. As a simple example, consider a scalar kinetic term $\frac{1}{2}\left(\partial_{\mu}\phi\right)^2$. In that case, $\Gamma_k^{(2)} = p^2$, and the trace will translate into an integral over all momenta. 

We will employ a truncation of the theory space where we keep the Newton coupling and cosmological constant, as well as wave-function renormalizations for all matter fields. We will neglect nonminimal matter-gravity couplings and matter interactions. Further, we will introduce a separate wave-function renormalization for the graviton\footnote{We use the term graviton to refer to the fluctuation field $h_{\mu \nu}$, which however is fully nonperturbative.}, that we will not identify with the Newton coupling, as in most previous truncations. For the case of pure gravity, a similar distinction has been made in \cite{Codello:2013fpa,Manrique:2009uh,Manrique:2010mq,Manrique:2010am,Donkin:2012ud,Christiansen:2012rx,Christiansen:2014raa}. 

Our truncation is given by an Einstein-Hilbert term,
\be
\Gamma_{\rm EH}= \frac{1}{16 \pi \bar{G}} \int d^dx \sqrt{g}\left(-R+2\bar{\lambda} \right).
\ee
As discussed above, it can be split into a background-dependent part $g_{\mu \nu} \rightarrow \bar{g}_{\mu \nu}$, from which we read off the running of the Newton coupling and cosmological constant. The graviton wave-function renormalization is then introduced as the prefactor of the quadratic term in the fluctuation field, after we redefine $h_{\mu \nu} \rightarrow \sqrt{32 \pi \bar{G}} h_{\mu \nu} $:
\bea
\delta^2 \Gamma_{\rm EH}&=&\frac{Z_h}{2} \int d^dx \sqrt{\bar{g}} 
\,h_{\mu\nu}K^{\mu\nu\alpha\beta} \cdot\nonumber\\
&{}& \cdot
((-\bar D^2-2\bar{\lambda})\mathbf{1}^{\rho\sigma}_{\alpha\beta}+W^{\rho\sigma}_{\alpha\beta})
h_{\rho\sigma} .
\nonumber
\eea
Here, $\mathbf{1}$ is the identity in the space of symmetric tensors and
$K_{\alpha\beta\rho\sigma}=\frac{1}{2}\left(\delta_{\alpha \rho}\delta_{\beta\sigma}
+\delta_{\alpha\sigma}\delta_{\beta\rho}
-\delta_{\alpha \beta} \delta_{\rho \sigma}\right)$.
The form of $W$ 
(see eq.(31) of \cite{Codello:2008vh}) is irrelevant for the calculation of the anomalous dimension, as we employ a flat background here.  Note that as an approximation (to be disentangled in the future) we will use the background couplings $\bar{G}$ and $\bar{\lambda}$ in the vertices and propagator for the graviton.

Further, the matter action takes the form
\bea
\Gamma_{\rm matter}&=& S_S+S_D+S_V 
\nonumber
\\
S_S&=&\frac{Z_S}{2} \int d^dx \sqrt{g}\,  g^{\mu \nu}\sum_{i=1}^{N_S}  \partial_{\mu} \phi^i \partial_{\nu} \phi^i
\nonumber\\
S_D&=& i Z_D \int d^dx \sqrt{g}\, \sum_{i=1}^{N_D} \bar{\psi}^i \slashed{\nabla} \psi^i,
\eea
with wave-function renormalizations $Z_S = Z_S(k)$ for the $N_S$ scalars and $Z_D= Z_D(k)$ for the $N_D$ fermions.
For the vector bosons with field strength $F_{\mu \nu} = \partial_{\mu}A_{\nu} - \partial_{\nu}A_{\mu}$, we have to include a gauge-fixing term, with gauge parameter $\xi$, and the corresponding ghost term:
\bea
S_V&=&
\frac{Z_V}{4}\int d^d x \sqrt{g}  \sum_{i=1}^{N_V}g^{\mu \nu}g^{\kappa \lambda}F^i_{\mu \kappa}F^i_{\nu \lambda}\nonumber\\
&{}& + \frac{Z_V}{2\xi} \int d^d x \sqrt{\bar{g}}\sum_{i=1}^{N_V}\left(\bar{g}^{\mu \nu}\bar{D}_{\mu}A^i_{\nu}\right)^2
\nonumber\\
&{}& + \frac{1}{2} \int d^d x \sqrt{\bar{g}}\sum_{i=1}^{N_V}\bar C_i(-\bar D^2)C_i\ .
\label{abelian_action}
\eea
Unlike abelian gauge theories in flat space, the ghost fields $C_i, \bar{C}_i$ cannot be neglected, since they couple into the flow of the background gravitational couplings.

All wave-function renormalizations $Z_{\Phi}$ ($\Phi= \{h, c, S, D, V \}$) are related to the corresponding anomalous dimensions $\eta_{\Phi}$ by $\eta_{\Phi} = - k \partial_k \ln Z_{\Phi}$.

In \cite{Dona:2012am} it was shown that only a particular choice of regulator function is compatible with the result obtained from a summation over the eigenvalues of the Dirac operator. In the following, we will use such a regulator, and choose a Litim-type shape function \cite{Litim:2001up}.
In \cite{Gies:2013noa}, it was shown that coupling fermions to quantum gravity does not necessarily require a formulation in terms of vielbeins, instead a translation into metric fluctuations can be used, as we will do here.

\subsection{The effect of spin-3/2 fields}
We will now add a spin-3/2 field, that should be understood as a gravitino, and therefore comes with the corresponding necessity to gauge-fix the local supersymmetry transformation. 
The action is given by
\be
S_{RS}=\frac{1}{2}\int d^4 x \sqrt{g} \epsilon^{\mu\nu\rho\sigma} \bar{\Psi}_\mu \gamma_\nu \gamma_5 \nabla_\rho \Psi_\sigma\ ,
\ee
We will employ a decomposition of the gravitino into irreducible components of SO(5), as we will work on a spherical background in four dimensions, such that $ \Psi_{\mu} = \psi_{\mu}^T + \left(\nabla - \frac{1}{4}\gamma_{\mu}\slashed{\nabla} \right) \chi + \frac{1}{4}\gamma_{\mu} \psi$,
where $\gamma^{\mu} \psi_{\mu}^T =0 = \nabla^{\mu} \psi_{\mu}^T$.

We follow \cite{Fradkin:1983mq}. The gauge-fixing is of the form
\bea
S_{\rm gf} &=&\frac{3}{16 \alpha'} \int d^4x \sqrt{g} \left(\alpha' \psi + \bar{\chi} \left(- \slashed{\nabla} +  i\sqrt{\frac{R}{3}}\right) \right)\cdot \nonumber\\
&{}& \cdot \left(\slashed{\nabla} +  i\sqrt{\frac{R}{3}} \right) \left(\alpha' \psi + \left(\slashed{\nabla} -  i\sqrt{\frac{R}{3}} \right)\chi \right).
\eea
We will set the gauge parameter $\alpha'=0$ in the following.
The presence of the operator $\slashed{\nabla} + i \sqrt{R/3}$ necessitates the introduction of additional ghosts, known as Nielsen-Kallosh ghosts \cite{Nielsen, Kallosh}. Their action reads as follows
\be
S_{NK}=\int d^4x \sqrt{g} \sum_i\bar{\omega}^{(i)}\left(\slashed{\nabla}+ i\sqrt{\frac{R}{3}}\right)\omega^{(i)},
\ee
where we have introduced a complex commuting Dirac spinor $ \omega^{(1)}$ and a real anticommuting Majorana spinor $\omega^{(2)}$.
The standard Faddeev-Popov ghost term for commuting Dirac spinors is
\be
S_{FP}= - \int d^4 x \sqrt{g} \bar{\eta} \left((\alpha'+1) \slashed{\nabla} - i\sqrt{\frac{R}{3}} \right) \eta.
\ee
In the presence of the ghosts, the counting of degrees of freedom for the gravitino works exactly: $\Psi_{\mu}$ is a Weyl spinor, with 8 degrees of freedom. The Faddeev-Popov ghosts subtract four, and the Nielsen-Kallosh ghosts subtract another 2, yielding the 2 propagating degrees of freedom of the gravitino.

We then choose so-called type-II cutoffs \cite{Codello:2008vh, Dona:2012am}, as used in \cite{Dona:2013qba} for all fields. The resulting contributions to the beta functions for $G$ and $\lambda$ are given in the next section.

\section{The effect of matter on the interacting gravitational fixed point}
\subsection{Perturbative analysis}
As a first step, let us consider a simplified truncation, where we disregard all wave-function renormalizations. We then make a transition to the dimensionless couplings defined by
\be
G= \bar{G} k^{d-2},\,\, \lambda=\bar{\lambda} k^{-2},
\ee
and then specialize to $d=4$.
Additionally, we will expand the $\beta$ functions for $G$ and $\lambda$ up to second order in $\lambda$ and $G$. They  take a simple form that is straightforward to analyze:
\bea
\beta_G &=& 2 G +\frac{G^2}{ 6 \pi} \left(N_S+2N_D-4 N_V - N_{RS} - 46 \right),\label{betaG}\\
\beta_{\lambda}&=& - 2 \lambda + \frac{G}{ 4 \pi} \left(N_S - 4N_D+2N_V-2 N_{RS}+2 \right) \nonumber\\
&{}&+ \frac{G \lambda}{6 \pi} \left(N_S +2N_D-4N_V - N_{RS}- 16\right).\label{betalambda}
\eea
The first term in each $\beta$ function comes from the canonical dimensionality of the couplings and would arise in a classical theory, where no quantum fluctuations introduce additional scale dependence. All other terms are induced by matter and gravity loops. Considering the matter-less case $N_V=N_S=N_D=N_{RS}=0$, we observe an interacting fixed point at $G_{\ast} =\frac{6 \pi}{23}$, $\lambda_{\ast}=\frac{3}{62}$. It lies at a positive value of the Newton coupling, as the term $\sim G^2$ in $\beta_{G}$ comes with a negative sign. Although the microscopic fixed-point value of the Newton coupling is of course not restricted by experiment, it is nevertheless required to be positive (in all known truncations), since $\beta_G (G=0)= 0$. In other words, starting with a negative value for $G$ in the ultraviolet, the RG flow stops at $G=0$ and can never cross over to the region $G>0$ where it has to end up in the infrared to be consistent with our universe. This observation immediately suggests that there will be a bound on the number of fermions and scalars that are consistent with asymptotic safety: Since the terms $\sim N_D$ and $\sim N_S$ in $\beta_G$ come with a positive sign, they can overwhelm the term arising from graviton fluctuations \footnote{Note that these are the same coefficients that can lead to a lowering of the ``quantum gravity scale", i.e., the scale at which $G \sim 1$, in the effective-field theory framework \cite{Calmet:2008df}.}. For $N_{ RS}=0$ and 
\be
N_S+2 N_D>46  + 4N_V,
\ee
the fixed point lies at $G_{\ast}<0$, which is ruled out from observations by the above argument. We can actually sharpen the bound, if we consider $\beta_{\lambda}$: The fixed-point value for $\lambda$ in the presence of matter is given by
\be
\lambda_{\ast} = - \frac{3}{4} \frac{N_S-4N_D+2N_V +2}{ N_S+2N_D - 4N_V -31}.
\ee
If we demand that the fixed point in the presence of matter is continuously connected to the pure-gravity one\footnote{While this requirement is not strictly speaking necessary, we will use it to distinguish fixed points which are likely to be truncation induced from fixed points which are likely to exist in full theory space: Since the pure-gravity fixed point has been found in numerous different truncations, it is most likely not a truncation artifact. Thus any continuously connected fixed point that is the deformation of the pure-gravity fixed point for nonvanishing matter, is most likely not an artifact. Further fixed points in the case with matter could in principle exist, but would have to be confirmed in extended truncations.}, then the singularity at
\be
N_S +2N_D = 4N_V +31,
\ee
yields a stronger upper bound on the allowed number of fermions and scalars.

We conclude that, within the truncation specified above, and for a fixed number of vectors, there is an upper bound on the number of fermions and scalars that are compatible with asymptotic safety.  The standard model has 12 vectors (1 photon, 8 gluons and 3 weak bosons), 4 scalars (one Higgs and three Goldstone modes that are ``eaten up" by the weak bosons and become their longitudinal degrees of freedom) and 45/2 Dirac fermions. The non-integer number of fermions arises, since the standard model is chiral, i.e., it is constructed from Weyl fermions. In the gravitational $\beta$ functions, a Dirac fermion essentially equals two Weyl fermions. As the standard model does not include right-handed neutrinos, it contains 45 Weyl fermions. These degrees of freedom are well-compatible with a viable gravitational fixed point in the simplified analysis based on eq.~\eqref{betaG} and eq.~\eqref{betalambda}.

We now turn to supersymmetric models, which must contain a gravitino.
As it contributes to the beta function for G
with the same sign as the graviton, it allows to extend the number
of matter fields slightly in comparison to the gravitino-less case.
In particular, simple SUGRA still admits a viable gravitational
fixed point.
The MSSM+SUGRA, however, contains too many fermions and
lies in the excluded region.

\subsection{Full analysis: Bounds on matter}
As a next step, we now include the effect of the wave-function renormalizations for gravitons, ghosts and matter fields, and do not perform a Taylor expansion in the couplings.
The $\beta$ functions for $G, \lambda$ then depend on the $\eta$'s and are considerably lengthier and can be found in \cite{Dona:2013qba}. 
The wave-function renormalizations play a special role: As they can be reabsorbed by a redefinition of the fields, there is no separate fixed-point requirement for those; they are called inessential couplings. In other words, the equations for $\eta_h, \eta_c$ and the anomalous dimensions of the matter fields can be solved in terms of $G, \lambda$. 
These can be inserted into $\beta_G$ and $\beta_{\lambda}$, which can then be used to determine the fixed-point coordinates $G_{\ast}, \lambda_{\ast}$. Once the fixed-point coordinates have been obtained, they can be reinserted into the expressions for the anomalous dimensions, and yield the fixed-point value of the anomalous dimensions. 
 Anomalous dimensions are crucial to determine the relevant operators at a fixed point. To see this, consider, e.g., an operator of the form $\phi^n$ with its coupling $\bar{g}_n$, which has canonical dimensionality $4-n$ in 4 spacetime dimensions. Rescaling the kinetic term to have canonical normalization $1/2$ and switching to dimensionless couplings yields $g_n = \bar{g}_n \frac{k^{n-4}}{Z_{\phi}^{n/2}}$. Thus the beta function for this coupling will contain the terms $\beta_{g_n} = (n-4) g_n + \frac{\eta_{\phi}}{2} g_n+...$. 
Accordingly, the critical exponents, which are determined by taking derivatives of the beta functions with respect to couplings, will receive contributions from the anomalous dimension. If we assume that the stability matrix Eq.~\ref{stabm} 
is diagonal, then the critical exponent corresponding to $g_n$ takes the form
\be
\theta_n = (4-n) - \frac{n}{2} \eta_{\phi}+...,
\ee
where the additional terms depend on the particular coupling under consideration.  A negative anomalous dimension shifts critical exponents towards positive values, implying that these couplings can become relevant. 
Accordingly it will be of interest to us to include the anomalous dimensions, not only because they affect the fixed-point equations, but also to see whether we should expect many relevant directions.

\begin{figure}[top]
\includegraphics[width=\linewidth]{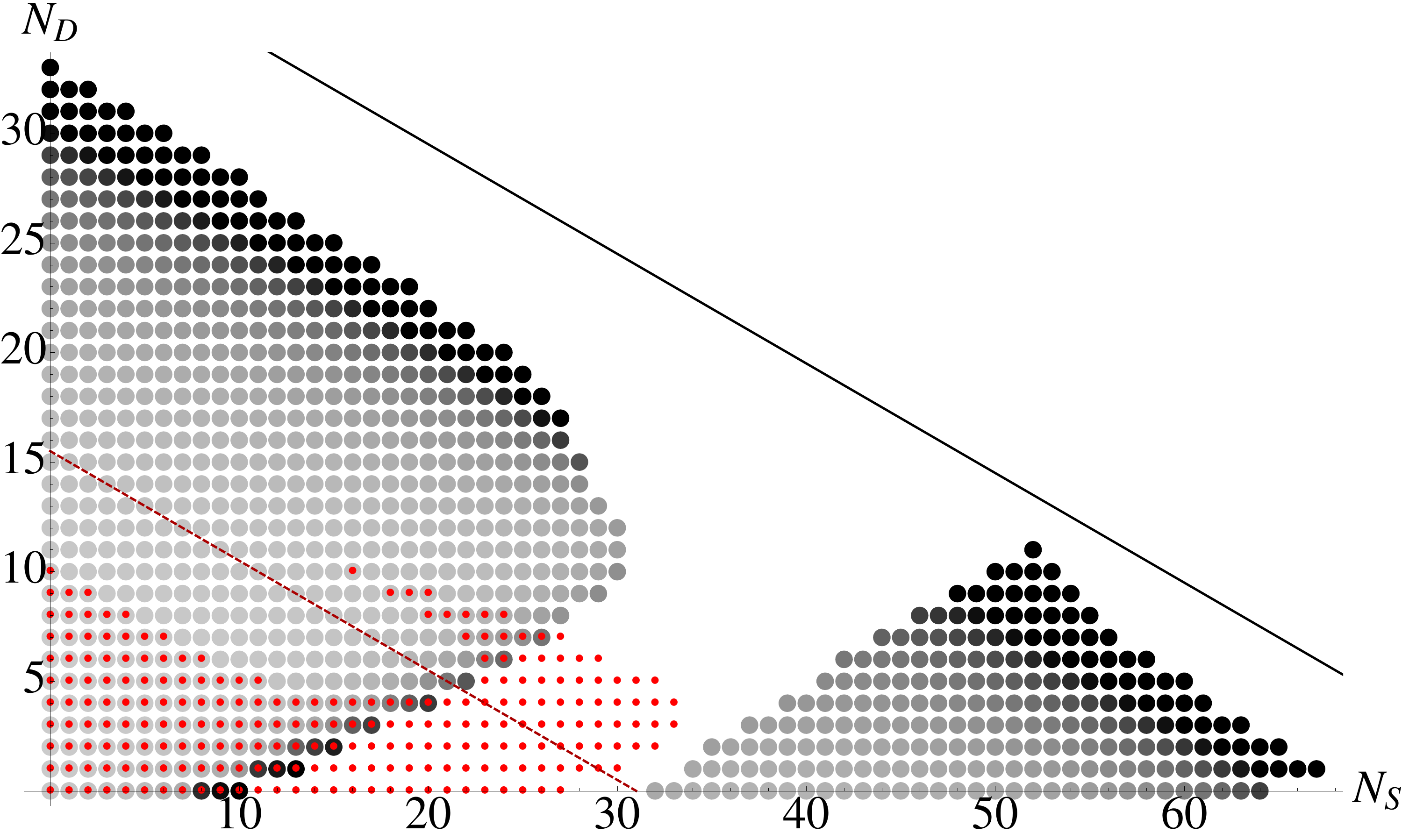}
\caption{\label{exclusionsplot} We plot the perturbative bound on the allowed number of fermions $N_D$ and scalars $N_S$ for $N_V=0$ (red dashed line) and $N_V=12$ (black thick line). Points which are compatible with a viable gravitational fixed point in the full analysis including wave function renormalizations are shown in red ($N_V=0$) and grey ($N_V=12$). The darker the grey, the larger the value of the graviton anomalous dimension, with black indicating $\eta_h>10$.}
\end{figure}

To find the interacting fixed point, we start at a vanishing number of matter fields, where we confirm results in \cite{Codello:2013fpa,Christiansen:2014raa}. We then increase the number of matter fields and numerically follow the fixed point. We impose the criterion that $G_{\ast}>0$ and that anomalous dimensions and critical exponents should stay bounded. The reason for the latter criterion lies in the fact that very large critical exponents and anomalous dimensions imply a very significant departure from canonical scaling, in which case our truncation is presumably insufficient, i.e., fixed points with these properties are most likely truncation artifacts. We also demand that the fixed point features two relevant directions. In this way, we again find bounds on the number of allowed fermions and scalars, if the number of vectors is fixed. As an example,  in fig.~\ref{exclusionsplot} we show the points $(N_S, N_D)$ with a viable gravitational fixed point for $N_V=0$ and $N_V=12$  and $N_{RS}=0$, with the data taken from \cite{Dona:2013qba}.

Although the shape of the boundary becomes more complicated, the qualitative results agree with our perturbative analysis.

Fermions shift the fixed-point value of the cosmological constant to negative values; an effect that can also be observed for the standard model. As the RG flow of $\lambda$ can cross the $\lambda=0$ plane towards the infrared, there is no problem with a negative fixed-point value and the observed small positive value in the infrared.

Most interestingly, we can now study matter theories which are of particular interest, which we report in tab.~\ref{FPtab}.\newline

\begin{table}
\begin{tabular}{cccccccc}
\rm{model} & $N_S$ & $N_D$ & $N_V$ & $G_{\ast}$ & $\lambda_{\ast}$ & $\theta_1$ & $\theta_2$ \\
\hline\hline
no matter & 0 & 0 & 0 & 0.77 & 0.01 & 3.30 & 1.95  \\ \hline
SM & 4 & 45/2 & 12 & 1.76 & -2.40 & 3.96 & 1.64 \\ \hline
SM + 3 $\nu$'s &4 & 24 & 12 & 2.15 & -3.20 & 3.97 & 1.65 \\ \hline
SM+3 $\nu$'s\\
+ 2 scalars & 6 & 24 & 12 & 2.50 & -3.61 & 3.96 & 1.63 \\ \hline
MSSM & 49 & 61/2 & 12 & x & x & x & x  \\ \hline
SU(5) GUT & 124 & 24 & 24 & x & x & x & x  \\ \hline
SO(10) GUT & 97 & 24 & 45 & x & x & x & x \\ \hline \hline
\end{tabular}
\begin{tabular}{cccccc}
\rm{model} & $\eta_h$ & $\eta_c$ & $\eta_S$ & $\eta_D$ & $\eta_V$\\
\hline\hline
no matter  & 0.27 & -0.81 & -- & -- & -- \\ \hline
SM  & 2.98 & -0.14 & -0.08 &-0.014 & -0.13 \\ \hline
SM + 3 $\nu$'s & 3.71 & -0.13 & -0.07 & -0.01 & -0.13\\ \hline
SM+3 $\nu$'s\\
+ 2 scalars &  4.28 & -0.13 & -0.07 & -0.004 & -0.14\\ \hline
MSSM & x & x & x& x& x \\ \hline
SU(5) GUT  & x & x & x& x& x \\ \hline
SO(10) GUT & x & x & x& x& x \\ \hline \hline
\end{tabular}
\caption{\label{FPtab} We report fixed-point values for various models, where SM stands for the standard model. The MSSM is the minimal supersymmetric standard model. The SU(5) and SO(10) GUT are candidates for Grand Unified Theories, which require a large number of scalars, such that even the increased number of vectors does not enable a viable gravitational fixed point.}
\end{table}

Again, the standard model lies well within the allowed region. This is a strong hint that, although extended truncations are likely to change the shape of the boundary, the standard model should remain within the allowed region also for extended truncations, if they are constructed consistently.

As discussed above, anomalous dimensions can shift couplings into (ir)relevance. It turns out that the graviton has a large positive anomalous dimension, shifting the corresponding operators very significantly towards irrelevance, and thus decreasing the number of free parameters, and making the theory more predictive. This effect becomes stronger with an increasing number of matter fields. Although the matter anomalous dimensions are all negative, they are rather small, and thus do not contribute to significant shifts of the critical exponents.

Observationally, we have good motivations to assume the existence of right-handed neutrinos, to allow for neutrino observations, adding 3/2 Dirac fermions to the matter content of the standard model, again leading to an allowed matter theory. Further observationally well-motivated theories consist of additional matter fields that constitute dark matter, and have to be coupled to gravity. In the simplest case \cite{Cline:2013gha,McDonald:1993ex,Burgess:2000yq,Eichhorn:2014qka}, a single additional scalar could constitute the complete dark matter relic density, and can easily be accommodated in the allowed region. Further single scalars can be introduced in the context of QCD in the form of an axion\footnote{Its pseudoscalar nature does not change its coupling to gravity at the minimal level.}, or in the context of cosmology.

On the other hand, a large number of scalars and fermions cannot be accommodated, unless a large number of vectors is postulated at the same time. Thus, the matter content of, e.g., the minimal supersymmetric standard model, or particular grand unified theories, is incompatible with a viable gravitational fixed point, at least within the truncation that we considered.

We conclude that only a restricted class of matter models seems to be compatible with a viable gravitational fixed point. This opens the possibility of experimentally testing asymptotically safe quantum gravity,  using data from the ongoing search for beyond-standard-model-particle physics scenarios, e.g., at the LHC. Clearly,  this should go hand-in-hand with a detailed study of extended truncations, taking into account further terms in the gravitational sector, as well as non minimal matter-gravity couplings and matter self interactions.

Our results suggest that a strategy, in which a consistent quantum theory of pure gravity is constructed first, and matter degrees of freedom are only added later, might actually not be successful, as many choices of matter theories seem to be inconsistent with asymptotically safe quantum gravity.

\subsection{Restrictions on extra dimensions}
Further, results we obtained in  \cite{Dona:2013qba} on scenarios with extra dimensions point to another possibility to test asymptotically safe quantum gravity: In settings with large extra dimensions, a fixed-point search in several truncations \cite{Fischer:2006fz,Ohta:2013uca} revealed a viable pure-gravity fixed point in $d \geq 4$. This would suggest that asymptotic safety is compatible with the existence of large extra dimensions, even though they are not required for the consistency of the theory. This could in principle open up the possibility for experimental tests of graviton-induced particle-scattering \cite{Litim:2007iu,Gerwick:2011jw,Dobrich:2012nv}. However, it turns out that a scenario with universal extra dimensions, where both gravity and matter propagate into the extra dimensions, is incompatible with the existence of the standard model for $d \geq 6$ within the above truncation. Generally, the allowed region in the $N_S, N_D$-plane for fixed $N_V$ shrinks rapidly, as the number of extra dimension is increased.
We conclude that an experimental discovery of universal extra dimension could pose a challenge for the asymptotic safety scenario. Note that scenarios in which only gravity can propagate into the extra dimensions are not restricted by these results.

\section{Summary}
Asymptotically safe quantum gravity is a model for a quantum field theory of gravity. Considerable evidence for the internal consistency of the model exists, so a next crucial step is to establish a link to observations and experiment. Clearly this is a highly nontrivial challenge. Here, we follow the route started in \cite{Dona:2013qba,Eichhorn:2011pc}, which aims at establishing experimental tests for quantum gravity by testing its consistency with observed  low-energy properties of matter.

The results reported here suggest that asymptotically safe quantum gravity could pass on important and nontrivial observational consistency test, as results within the truncation  in \cite{Dona:2013qba} indicate that the standard model matter content is compatible with a viable gravitational fixed point. Studies in extended truncations will allow us to confirm these indications. The limitations of our calculation are obvious: While the ``bimetric" structure of gravitational RG flows has been considered, higher-order vertex functions for the graviton could play an important role. Furthermore, non minimal matter-curvature couplings have been ignored and could become important.

In this work, we have added the contribution of spin-3/2 gravitinos. We find that a model of pure supergravity admits an asymptotically safe gravitational fixed point. If we add matter, bounds on the number of allowed matter fields persists. In particular, the matter content of the MSSM is not compatible with a viable fixed point within a model with one graviton and one gravitino.

Keeping these limitations of the current work in mind, we conclude that it could be possible to experimentally test asymptotically safe quantum gravity, at present and future particle colliders and other particle searches: If models such as, e.g., particular supersymmetric models, were confirmed experimentally, the results presented here would suggest that there is no viable gravitational fixed point. (In such a case it might be crucial to include supersymmetric interaction terms in extended truncations and use regulators which respect supersymmetry \cite{Synatschke:2008pv}.)

This provides a possible experimental test of asymptotically safe quantum gravity, and might allow significant progress in the search for a consistent model of quantum spacetime. 

\emph{Acknowledgements:} This research was supported in part by Perimeter Institute for Theoretical Physics. Research at Perimeter Institute is supported by the Government of Canada
through Industry Canada and by the Province of Ontario through the Ministry of Research and Innovation.

\end{document}